\documentclass[10pt, conference, compsocconf]{IEEEtran}
\ifCLASSINFOpdf
\else
\fi

\usepackage{comment}

\usepackage{mathtools}
\usepackage{amssymb}
\usepackage{amsthm}
\usepackage{tabularx}
\usepackage[makeroom]{cancel}
\usepackage{multirow}

\usepackage{breqn}
\usepackage{enumitem} 
\usepackage{mdwlist}

\usepackage{hyperref}
\newcommand{\LINK}{{\color{blue} \href{http://sybrandt.com/2018/validation}{sybrandt.com/2018/validation}}}

\newcommand{\RR}{\mathbb{R}}

\newcommand{\net}{\mathcal{N}}

\usepackage{xspace}
\newcommand{\semmeddb}{\textsf{SemMedDB}\xspace}
\newcommand{\medline}{\textsf{MEDLINE}\xspace}

\newcommand{\umls}{\textsf{UMLS}\xspace}
\newcommand{\ftext}{\textsf{FastText}\xspace}
\newcommand{\sysname}{\textsf{MOLIERE}\xspace}
\newcommand{\tmine}{\textsf{ToPMine}\xspace}
\newcommand{\wvec}{\textsf{word2vec}\xspace}
\newcommand{\arrowsmith}{\textsf{Arrowsmith}\xspace}

\newcommand{\csim}{\textsc{CSim}}

\usepackage{changes}
\definechangesauthor[color=red]{is}
\definechangesauthor[color=blue]{js}

\usepackage{graphicx}
\usepackage{calc}
\newlength{\depthofsumsign}
\newlength{\totalheightofsumsign}
\newlength{\heightanddepthofargument}

\makeatletter
\newcommand*{\DivideLengths}[2]{%
  \strip@pt\dimexpr\number\numexpr\number\dimexpr#1\relax*65536/\number\dimexpr#2\relax\relax sp\relax
}
\makeatother

\graphicspath{{img/}}

\usepackage{algpseudocode}
\usepackage{algorithm}

\algblock{ParFor}{EndParFor}
\algnewcommand\algorithmicparfor{\textbf{for}}
\algnewcommand\algorithmicpardo{\textbf{pardo}}
\algnewcommand\algorithmicendparfor{\textbf{end\ for}}
\algrenewtext{ParFor}[1]{\algorithmicparfor\ #1\ \algorithmicpardo}
\algrenewtext{EndParFor}{\algorithmicendparfor}

\begin{document}
%

\title{Large-Scale Validation of Hypothesis Generation Systems via Candidate Ranking}


\author{\IEEEauthorblockN{
Justin Sybrandt
}
\IEEEauthorblockA{
Clemson University\\
School of Computing\\
Clemson, USA\\
jsybran@clemson.edu}
\and
\IEEEauthorblockN{
Michael Shtutman
}
\IEEEauthorblockA{
University of South Carolina\\
Drug Discovery and Biomedical Sciences\\
Columbia, USA\\
shtutmanm@sccp.sc.edu}
\and
\IEEEauthorblockN{
Ilya Safro
}
\IEEEauthorblockA{
Clemson University\\
School of Computing\\
Clemson, USA\\
isafro@clemson.edu}
}


%


\maketitle

\begin{abstract}

The first step of many research projects is to define and rank a short list of candidates for study.
In the modern rapidity of scientific progress, some turn to automated hypothesis generation (HG) systems to aid this process.
These systems can identify implicit or overlooked connections within a large scientific corpus, and while their importance grows alongside the pace of science, they lack thorough validation.
Without any standard numerical evaluation method, many validate general-purpose HG systems by rediscovering a handful of historical findings, and some wishing to be more thorough may run laboratory experiments based on automatic suggestions.
These methods are expensive, time consuming, and cannot scale.
Thus, we present a numerical evaluation framework for the purpose of validating HG systems that leverages thousands of validation hypotheses.
This method evaluates a HG system by its ability to rank hypotheses by plausibility; a process reminiscent of human candidate selection.
Because HG systems do not produce a ranking criteria, specifically those that produce topic models, we additionally present novel metrics to quantify the plausibility of hypotheses given topic model system output.
Finally, we demonstrate that our proposed validation method aligns with real-world research goals by deploying our method within \sysname, our recent topic-driven HG system, in order to automatically generate a set of candidate genes related to HIV-associated neurodegenerative disease (HAND).
By performing laboratory experiments based on this candidate set, we discover a new connection between HAND and Dead Box RNA Helicase 3 (DDX3).\\
Reproducibility: code, validation data, and results can be found at \LINK.
\end{abstract}

%
\IEEEpeerreviewmaketitle

\section{Introduction}\label{sec:introduction}

In the early stages of a research project, biomedical scientists often perform ``candidate selection,'' wherein they select potential targets for future study~\cite{jekunen2014decision}.
For instance, when exploring a certain cancer, scientists may identify a few dozen genes on which to experiment.
This process relies on the background knowledge and intuitions held by each researcher, and  higher-quality candidate lists often lead to more efficient research results.
However, the rate of scientific progress has been increasing steadily~\cite{van2014global}, and occasionally scientists miss important findings.
for instance, was the case regarding the missing connection between Raynaud's Syndrome and fish oil~\cite{swanson1986fish}, and in the case of five genes recently linked to Amyotrophic Lateral Sclerosis~\cite{bakkar2018artificial}.
Hypothesis Generation (HG) systems allow scientists to leverage the cumulative knowledge contained across millions of papers, which lead to both above findings, among many others.
The importance of these systems rises alongside the pace of scientific output; an abundance of literature implies an abundance of overlooked connections.
While many propose techniques to understand potential connections~\cite{wang2011finding, sybrandt2017moliere, liu2014diseaseconnect, swanson1986undiscovered, spangler2015accelerating}, few \emph{automated} validation techniques exist~\cite{bruza2008literature} for general-purpose HG systems (not designed for specific sub-domains or types of queries such as OHSUMED  \cite{hersh1994ohsumed} or BioCreative datasets).
Often, subject-matter experts assist in validation by running laboratory experiments based on HG system output.
This process is expensive, time consuming, and does not scale beyond a handful of validation examples.

HG systems are hard to validate because they attempt to uncover novel information, unknown to even those constructing or testing the system.
For instance, how are we to distinguish a bizarre generated hypothesis that turns out to produce important results from one that turns out to be incorrect?
Furthermore, how can we do so at scale or across fields?
While there are verifiable models for novelty in specific contexts, each is trained to detect patterns similar to those present in a training set, which is conducive to traditional cross-validation.
Some examples include using non-negative matrix factorization to uncover protein-protein interactions~\cite{greene2008ensemble}, or to discover mutational cancer signatures~\cite{alexandrov2013signatures}.
However, HG is unlike the above examples as it strives to detect novel patterns that are a) \emph{absent} from a dataset, b) may be wholly unknown or even currently counterintuitive, and c) not necessarily outliers as in traditional data mining.

{\bf Our contribution:}
In this paper we propose novel hypothesis ranking methods and a method to validate HG systems that does not require expert input and allows for large validation sets.
This method judges a system by its ability to rank hypotheses by plausibility, similarly to how a human scientist must rank potential research directions during candidate selection.
We start by dividing a corpus based on a ``cut date,'' and provide a system only information that was priorly available.
Then, we identify predicates (clauses consisting of subject, verb, and object) whose first co-occurrence in a sentence is after the cut date.
Because typical corpora contain only titles and abstracts, these recently introduced connections represent significant findings that were not previously formulated, thus we can treat them as surrogates for plausible hypotheses from the perspective of the system under evaluation.
To provide implausible hypotheses, we randomly generate predicates that do not occur in the corpus as a whole.
Then, the HG system must rank both the plausible and implausible predicates together by evaluating the predicted connection strength between each predicate's subject and object.
The system's evaluation is based on the area under this ranking's Receiver Operating Characteristic (ROC) curve, wherein the highest area under curve (AUC) of 1 represents a ranking that places all plausible connections above the implausible, and the lowest AUC of 0.5 represents an even mixture of the two.

We note that many HG systems do not typically produce a ranking criteria for potential hypotheses.
Particularly, we find that those systems that produce topic model output, such as \sysname~\cite{sybrandt2017moliere} or BioLDA~\cite{wang2011finding}, lack this criteria, but present promising results through expert analysis.
Therefore, we additionally developed a number of novel metrics for topic-driven HG systems that quantify the plausibility of potential connections.
These metrics leverage word embeddings~\cite{mikolov2013efficient} to understand how the elements of a hypothesis relate to its resulting LDA topic model~\cite{blei2003latent}.
Through our experiments, described below, we identify that a polynomial combination of five different metrics allows for the highest-scoring ranking (0.834).
This result is especially significant given that the main validation methods available, to both \sysname and other similar systems (see survey in \cite{sybrandt2017moliere}), were expert analysis and replicating the results of others~\cite{bruza2008literature}.
Still, while the systems mentioned above focus on the medical domain, we note that neither our metrics, nor our validation methodology, are domain specific.

To demonstrate that our proposed validation process and new metrics apply to real-world applications, we present a case study wherein our techniques validate an open-source HG system as well as identify a novel gene-disease connection.
We modify \sysname to support our new metrics, and we perform our validation process.
This system is trained on \medline~\cite{Medicine2016}, a database containing over 27 million papers (titles and abstracts) maintained by the National Library of Health.
We use \semmeddb~\cite{kilicoglu2012semmeddb}, a database of predicates extracted from \medline, in order to identify the set of ``published'' (plausible) and ``noise'' (implausible) hypotheses.
This database represents its connections in terms of codified entities provided by the Unified Medical Language System (\umls), which enables our experimental procedure to be both reproducible and directly applicable to many other medical HG systems.
This evaluation results in an ROC AUC of 0.834, and when limiting the published set to only predicates occurring in papers that received over 100 citations, this rises to 0.874.
Then, we generate hypotheses, using up-to-date training data, which attempt to connect HIV-associated neurodegenerative disease (HAND) to over 30,000 human genes.
From there, we select the top 1,000 genes based on our ranking metrics as a large and rudimentary ``candidate set.''
By performing laboratory experiments on select genes within our automatically generated set, we discover a new relation between HAND and Dead Box RNA Helicasee 3 (DDX3).
Thus, demonstrating the practical utility of our proposed validation and ranking method.

\section{Technical Background}\label{sec:background}

\noindent {\bf Extracting Information from Hypothesis Generation Systems}
Swanson and Smalheiser created the first  HG system \arrowsmith~\cite{smalheiser1998using}, and in doing so outlined the ABC model for discovery~\cite{swanson1997interactive}.
Although this approach has limitations~\cite{smalheiser2012literature}, its conventions and intuitions remain in modern approaches~\cite{spangler2015accelerating}. 

In the ABC model, users run queries by specifying two keywords $a$ and $c$.
From there, the goal of a  HG system is to discover some entity $b$ such that there are known relationships ``$a \rightarrow b$'' and ``$b \rightarrow c$,'' which allow us to infer the relationship between $a$ and $c$.
Because many connections may require more than one element $b$ to describe, researchers apply other techniques, such as topic models in our case, to describe these connections.

We center this work around the \sysname HG system~\cite{sybrandt2017moliere}.
Once a user queries $a$ and $c$, the system identifies a relevant region within its multi-layered knowledge network, which consists of papers, terms, phrases, and various types of links.
The system then extracts abstracts and titles from this region and creates a sub-corpus upon which we generate a topic model
(Note that in~\cite{sybrandt2018abstracts} we address trade-offs of using full text).
This topic model describes groups of related terms, which we study to understand the quality of the $a$-to-$c$ connection.
Previously, these results were compared biased on those words that co-occur with high probability in prominent topics.
Without clear metrics, or a validation framework, experts could only help evaluate a select handful of $a$, $c$ pairs.

\smallskip

\noindent {\bf Word and Phrase Embedding} 
The method of finding dense vector representations of words is often referred to as ``\wvec.''
In reality, this umbrella term references two different algorithms, the Continuous BOW (CBOW) method and the Skip-Gram method~\cite{mikolov2013efficient}. 
Both rely on shallow neural networks in order to learn vectors through word-usage patterns.

\begin{figure}
\centering
\includegraphics[width=.8\linewidth]{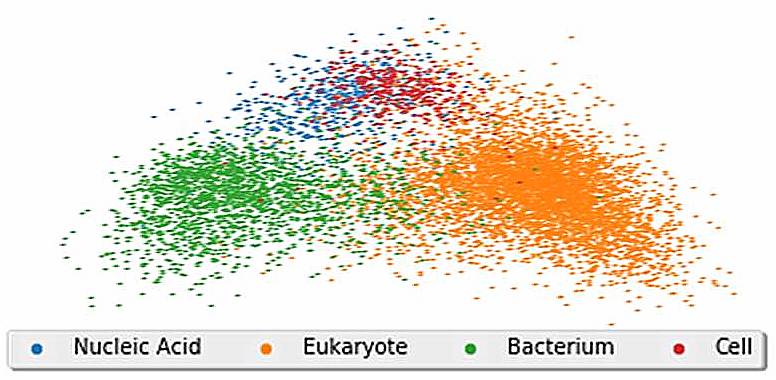}
\caption{
The above diagram shows a 2-D representation of the embeddings for over 8 thousand UMLS keywords within MOLIERE.
We used singular value decomposition to reduce the dimensionality of these vectors from 500 to 2.
}
\label{fig:word2vec_embeddingspace}
\end{figure}

\sysname uses \ftext~\cite{joulin2016fasttext}, a similar tool under the \wvec umbrella, to find high-quality embeddings of medical entities.
By preprocessing \medline text with the automatic phrase mining technique \tmine~\cite{el2014scalable}, we improve these embeddings while finding multi-word medical terms such as ``lung cancer'' or ``benign tumor.''
We see in Figure~\ref{fig:word2vec_embeddingspace} that \ftext clusters similar biological terms, an observation we later leverage to derive a number of metrics.

\smallskip

\noindent {\bf Topic Models}
Latent Dirichlet Allocation (LDA)~\cite{blei2003latent}, the classical topic modeling method, groups keywords based on their document co-occurrence rates in order to describe the set of trends that are expressed across a corpus.
A topic is simply a probability distribution over a vocabulary, and each document from the input corpus is assumed to be a mixture of these topics.
For instance, a topic model derived from New York Times articles would likely find one topic containing words such as ``computer,'' ``website,'' and ``Internet,'' while another topic may contain words such as ``money,'' ``market,'' and ``stock.''

In the medical domain, some use topic models to understand trends across scientific literature.
We look for groupings of entities such as genes, drugs, and diseases, which we then analyze to find novel connections.
While LDA is the classical algorithm, \sysname uses a parallel technique, PLDA+~\cite{liu2011plda+} to quickly find topics from documents related to $a$ and $c$.
Additionally, because \sysname preprocess's \medline articles with \tmine, its resulting topic models include both words and phrases.
This often leads to more interpretable results, as a topic containing an n-gram, such as ``smoking induced asthma,'' is typically easier to understand than a topic containing each unigram listed separately with different probabilities.

We additionally can use the probabilities of each word to represent a topic within an embedding space created with \wvec.
For instance, we can take a weighted average over the embeddings for each topic to describe each topics's ``center.''
Additionally, we can simply treat each topic as a weighted point cloud for the purposes of typical similarity metrics.
We leverage both representations later in our metrics.

\section{Validation Methodology}\label{sec:validationChallenge}

In order to unyoke automatic  HG from expert analysis, we propose a method that any system can leverage, provided it can rank its proposed connections.
A successful system ought to rank published connections higher than those we  randomly created.
We train a system given historical information, and create the ``published,'' ``highly-cited,'' and ``noise'' query sets.
We pose these connections to an HG system, and rank its outputs in order to plot ROC curves, which determine whether published predicates are preferred to noise.
Through the area under these ROC curves, a  HG system demonstrates its quality at a large scale without expert analysis.

Our challenge starts with the Semantic Medical Database (\semmeddb)~\cite{kilicoglu2012semmeddb} that contains predicates extracted from \medline defined on the set of \umls terms~\cite{Medicine2016}.
For instance, predicate ``C1619966 TREATS C0041296'' represents a discovered fact ``abatacept treats tuberculosis.''
Because \sysname does not account for word order or verb, we look for distinct unordered word-pairs $a$--$c$ instead.
In Section~\ref{sec:deploymentChallenges}, we discuss how we may improve \sysname to include this unused information.

From there, we select a ``cut year.''
Using the metadata associated with each predicate, we note the date each unordered pair was first published.
For this challenge, we train \sysname using only information published before the cut year.
We then identify the set of \semmeddb unordered pairs $a$--$c$ first published after the cut year provided $a$ and $c$ both occur in that year's \umls release.
This ``published set'' of pairs represent new connections between existing entities, from the perspective of the HG system.
We select 2010 as the cut year for our study in order to create a published set of over 1 million pairs.
(Due to practical limitations, our evaluation consists of a randomly chosen subset of 4,319 pairs.)

Additionally, we create a set of ``highly-cited'' pairs by filtering the published set by citation count.
We use data from \semmeddb, \medline, and Semantic Scholar to identify 1,448 pairs from the published set that first occur in a paper cited over 100 times.
We note that this set is closer to the number of landmark discoveries since the cut-date, given that the published set is large and likely contains incidental or incorrect connections.

To provide negative examples, we generate a ``noise set'' of pairs by sampling the cut-year's \umls release, storing the pair only if it does not occur in \semmeddb.
These pairs represent nonsensical connections between \umls elements.
Although it is possible that we may stumble across novel findings within the noise set, we assume this will occur infrequently enough to not affect our results.
We generate two noise pair sets of equal size to both the published and highly-cited sets.

We run $a$--$c$ queries from each set through \sysname and create two ranked lists: published vs. noise (PvN) (8,638 total pairs) and highly-cited vs. noise (HCvN) (2,896 total pairs).
After ranking each set, we generate ROC curves \cite{han2011data}, which allow us to judge the quality of an HG system.
If more published predicates occur earlier in the ranking than noise, the ROC area will be close to 1; otherwise it will be closer to 0.5.



\section{New Ranking Methods for Topic Model Driven Hypotheses}\label{sec:methodology}

Because many HG systems do not currently produce a ranking criteria, such as those systems that instead return topic models~\cite{sybrandt2017moliere,wang2011finding}, we propose here a number of metrics to numerically evaluate the plausibility of potential connections.
We implement these metrics within \sysname~\cite{sybrandt2017moliere}.
This system is open source, and already leverages word embeddings in order to produce topic model output for potential connections --- all of which are properties our metrics exploit.
Put simply, \sysname takes as input two keywords ($a$ and $c$), and produces a topic model ($T$) that describes the structure of relevant documents.

While these metrics are proposed in the context of validation, another extremely important use case is that of the \emph{one-to-many} query.
Often during candidate selection, scientists may have a large list of initial potential targets --- such as 30,000 genes in the human genome --- that they wish to consider.
For this, one may run a large set of queries between some disease $a$, and each target $c_i$.
However, without a ranking criteria, the analysis of each $a$--$c_i$ connection is left to experts, which is untenable for most practical purposes.

\begin{figure}
\centering
\includegraphics[width=0.5\linewidth]{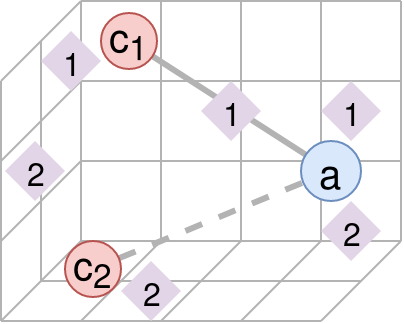}
\caption{\label{fig:topicsInVectorSpace}
The above depicts two queries, $a$--$c_1$ and $a$--$c_2$, where $a$--$c_1$ is a published connection and $a$--$c_2$ is a noise connection. 
We see topics for each query represented as diamonds via $\textsc{Centr}(T_i)$.
Although both queries lead to topics which are similar to $a$, $c_1$, or $c_2$, we find that the the presence of some topic which is similar to \emph{both} objects of interest may indicate the published connection.
}
\end{figure}

To begin, we note a key intuition underpinning the following metrics, depicted in Figure \ref{fig:topicsInVectorSpace}.
Not only are related objects grouped in a word embedding space, but the distances between words are also meaningful.
For this reason we hypothesize, and later show through validation experiments, that one can estimate the strength of an $a$--$c$ connection by comparing the distance of topics to the embeddings of each $a$, $c$, and their midpoint.
Note, we use $\epsilon(x)$ to map a text object $x$ into this embedding space, as described in~\cite{mikolov2013efficient}.
But, because not all hypotheses or topic models exhibit the same features, we quantify this ``closeness'' in eleven ways, and then train a polynomial to weight the relevance of each proposed metric.

\subsection{Similarity Between Query Words}\label{sec:similarityBetweenQueryWords}

As a baseline, we first consider two similarity metrics that do not include topic information: cosine similarity ($\csim$) and Euclidean distance ($L_2$):
\begin{equation*}
\csim(a,c) = \frac{\epsilon(a) \cdot \epsilon(c)}
                {||\epsilon(a)||_2 \times ||\epsilon(c)||_2}
\text{ , }
L_2(a,c) = ||\epsilon(a) - \epsilon(c)||_2,
\end{equation*}
where $a$ and $c$ are the two objects of interest, and $\epsilon(x)$ is an embedding function (see Section \ref{sec:background}).
Note that when calculating ROC curves for the $L_2$ metric, we will sort in reverse, meaning smaller distances ought to indicate published predicates.

These metrics indicate whether $a$ and $c$ share the same cluster with respect to the embedding space.
Our observation is that this can be a good indication that $a$ and $c$ are of the same kind, or are conceptually related. 
This cluster intuition is shared by others studying similar embedding spaces~\cite{wang2016semantic}.

\subsection{Topic Model Correlation}\label{sec:topicModelCorrelation}

The next metric attempts to uncover whether $a$ and $c$ are mutually similar to the generated topic model. 
This metric starts by creating vectors $v(a,T)$ and $v(c,T)$ which express each object's similarity to topic model $T = \{T_i\}_{i=1}^k$ derived from an $a-c$ query. 
We do so by calculating the weighted cosine similarity $\textsc{TopicSim}(x, T_i)$ between each topic $T_i$ and each object $x\in \{a,c\}$, namely,
\begin{equation*}
\textsc{TopSim}(x, T_i) = \sum_{(w, p) \in T_i} p \cdot \csim(x, w),
\end{equation*}
where a probability distribution over terms in $T_i$ is represented as word-probability pairs $(w, p)$. This metric results in a value in the interval [-1, 1] to represent the weighted similarity of $x$ with $T_i$. 
The final similarity vectors $v(a,T)$ and $v(c,T)$ in $\RR^k$ are defined below.
\begin{equation*}
\forall x\in \{a,c\} ~~~ v(x, T) = \begin{bmatrix}
\textsc{TopSim}(x, T_1) \\
\textsc{TopSim}(x, T_2) \\
\vdots\\
\textsc{TopSim}(x, T_k) \\
\end{bmatrix}
\end{equation*}



Finally, we can see how well $T$ correlates with both $a$ and $c$ by  taking another cosine similarity

\begin{equation*}
\textsc{TopicCorr}(a, c, T) = \frac{v(a, T) \cdot v(c, T)}
                                         {||v(a, T)||_2 \times  ||v(c, T)||_2}\in [-1,1].
\end{equation*}

If $\textsc{TopicCorr}(a, c, T)$ is close to 1, then topics that are similar or dissimilar to $a$ are also similar or dissimilar to $c$.
Our preliminary results show that if some explanation of the $a-c$ connection exists within $T$, then many $T_i\in T$ will likely share these similarity relationships.

\subsection{Similarity of Best Topic Centroid}\label{sec:similarityOfBestTopicCentroid}

While the above metric attempts to find a trend within the entire topic model $T$, this metric attempts to find just a single topic $T_i\in T$ that is likely to explain the $a-c$ connection.
This metric is most similar to that depicted in Figure \ref{fig:topicsInVectorSpace}.
Each $T_i$ is represented in the embedding space by taking a weighted centroid over its word probability distribution.
We then rate each topic by averaging its similarity with both queried words.
The score for the overall hypothesis is simply the highest score among the topics.

We define the centroid of $T_i$ as
\begin{equation*}
\textsc{Centr}(T_i) = \sum_{(w,p) \in T_i} \epsilon(w) \cdot p,
\end{equation*}
and then compare it to both $a$ and $c$ through cosine similarity and Euclidean distance.
When comparing with $\csim$, we highly rank $T_i$'s with centroids located within the arc between $\epsilon(a)$ and $\epsilon(c)$.
Because our embedding space identifies dimensions that help distinguish different types of objects, and because we trained a 500-dimensional embedding space, cosine similarity intuitively finds topics that share similar characteristics to both objects of interests.
We define the best centroid similarity for $\csim$ as 
\begin{equation*}
\textsc{BestCentrCSim}(a, c, T) = 
\max_{T_i \in T} \frac{\csim(a, T_i) + \csim(c, T_i)}{2}.
\end{equation*}

What we lose in the cosine similarity formulation is that clusters within our embedding space may be separate with respect to Euclidean distance but not cosine similarity.
In order to evaluate the effect of this observation, we also formulate the best centroid metric with $L_2$ distance.
In this formulation we look for topics that occur as close to the midpoint between $\epsilon(a)$ and $\epsilon(c)$ as possible.
We express this score as a ratio between that distance and the radius of the sphere with diameter from $\epsilon(a)$ to $\epsilon(c)$.
In order to keep this metric in a similar range to the others, we limit its range to [0, 1], namely, for the midpoint $m = (\epsilon(a) + \epsilon(c))/2$.
\begin{equation*}
\textsc{BestCentrL}_2(a, c, T) = 
\max_{T_i \in T}\left\lbrace
1 - \frac{\left\lVert \textsc{Centr}(T_i) - m\right\rVert_2}
         {\left\lVert m \right\rVert_2}\right\rbrace
\end{equation*}

\subsection{Cosine Similarly of Best Topic Per Word}\label{sec:cosineSimilarityOfBestTopicPerWord}

In a similar effort to the above centroid-based metric, we attempt to find topics which are related to $a$ and $c$, but this time on a per-word (or phrase) basis using $\textsc{TopicSim}(x, T_i)$ from Section~\ref{sec:topicModelCorrelation}.
Now instead of looking across the entire topic model, we attempt to identify a single topic which is similar to both objects of interest.
We do so by rating each topic by the lower of its two similarities, meaning the best topic overall will be similar to both query words.

\begin{equation*}
\textsc{BestTopPerWord}(a, c, T) = \max_{T_i \in T} \min
\left(\begin{matrix}
\textsc{TopSim}(a, T_i), \\
\textsc{TopSim}(c, T_i)
\end{matrix} \right)
\end{equation*}

\subsection{Network of Topic Centroids}
\label{sec:topicWalk}

A majority of the above metrics rely on a single topic to describe the potential connection between $a$ and $c$, but as Smalheizer points out in~\cite{smalheiser2017rediscovering}, a hypothesis may be best described as a ``story'' --- a series of topics in our case.
To model semantic connections between topics, we induce a nearest-neighbors network $\net$ from the set of vectors $V=\epsilon(a) \cup \epsilon(b) \cup \{\textsc{Centr}(T_i) | T_i \in T\}$ which form the set of nodes for $\net$.
In this case, we set the number of neighbors per node to the smallest value (that may be different for each query) such that there exists a path from $a$ to $c$.
Using this topic network, we attempt to model the semantic differences between published and noise predicates using network analytic metrics.

\begin{figure}
\includegraphics[width=\linewidth]{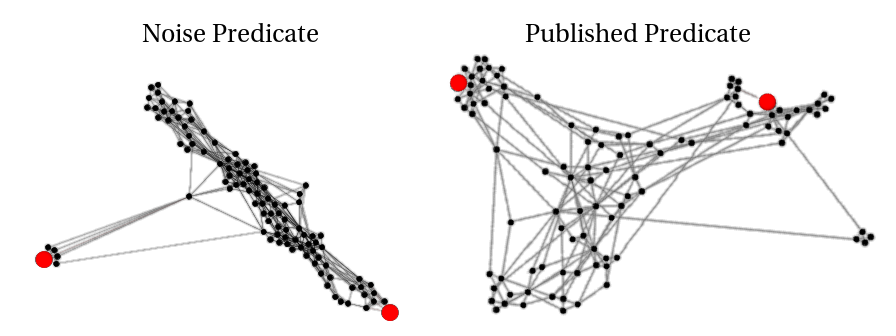}
\caption{\label{fig:topicGraph}
Above depicts two topic networks as described in Section~\ref{sec:topicWalk}.
In this visualization, longer edges correspond to dissimilar neighbors.
In red are objects $a$ and $c$, which we queried to create these topic models.
We observe that the connectivity between $a$ and $c$ from the published predicate is much higher than in the noisy example.
}
\end{figure}

We depict two such networks in Figure~\ref{fig:topicGraph}, and observe that the connectivity between $a$ and $c$ from a published predicate is substantially stronger and more structured.
In order to quantify this observed difference, we measure the average betweenness and eigenvector centrality~\cite{newman2010networks} of nodes along a shortest path from $a$ to $c$ (denoted by $a\sim c$) within $\net$ to reflect possible information flow between $T_i\in T$.
This shortest path represents the series of links between key concepts present within our dataset that one might use to explain the relationship between $a$ and $c$.
We expect the connection linking $a$ and $c$ to be stronger if that path is more central to the topic network.
Below we define metrics to quantify the differences in these topic networks. Such network analytic metrics are widely applied in semantic knowledge networks \cite{sowa2014principles}.

\noindent\textsc{TopWalkLength}$(a, c, T)$:~Length of shortest path $a \sim c$\\
\noindent\textsc{TopWalkBtwn}$(a, c, T)$:~Avg. $a\sim c$ betweenness centrality\\
\noindent\textsc{TopWalkEigen}$(a, c, T)$:~Avg. $a\sim c$ eigenvalue centrality\\
\noindent\textsc{TopNetCCoef}$(a, c, T)$:~Clustering coefficient of $\net$\\
\noindent\textsc{TopNetMod}$(a, c, T)$:~Modularity of $\net$\\

\subsection{Combination of Multiple Metrics}\label{sec:combinationOfMultipleMetrics}
Each of the above methods are based on different assumptions regarding topic model or embedding space properties exhibited by published connections.
To leverage each metric's strengths, we combined the top performing ones from each category into the following \textsc{PolyMultiple} method. 
We explored polynomial combinations in the form of $\sum_i \alpha_i x_i ^{\beta_i}$ for ranges of $\alpha_i \in [-1, 1]$ and $\beta_i \in [1, 3]$ after scaling each $x_i$ to the $[0, 1]$ interval.
Through a blackbox optimization technique, we searched over one-million parameter combinations.
In doing so we maximize for the AUC of our validation curve by sampling each $\alpha_i$ and $\beta_i$ from their respective domains.
We perform this search stochastically, sampling from parameter space and limiting our search space as we find stable local-minima.
Our results represent the best parameter values determined after one-million parameter samples.

\begin{multline*}
\textsc{PolyMultiple}(a, c, T) =
 \alpha_1 \cdot L_2^{\beta_1} + \alpha_2 \cdot \textsc{BestCenterL}_2^{\beta_2} \\
 +\alpha_3 \cdot \textsc{BestTopPerWord}(a, c, T)^{\beta_3} + \alpha_4 \cdot \textsc{TopCorr}(a, c, T)^{\beta_4} \\
 \alpha_5 \cdot \textsc{TopWalkBtwn}(a, c, T)^{\beta_5} +  \alpha_6 \cdot \textsc{TopNetCCoef}(a, c, T)^{\beta_6}
\end{multline*}

\section{Results and Lessons Learned}\label{sec:results}

\begin{figure*}
\centering
\begin{minipage}[b]{0.45\textwidth}
\centering
\includegraphics[width=0.8\linewidth]{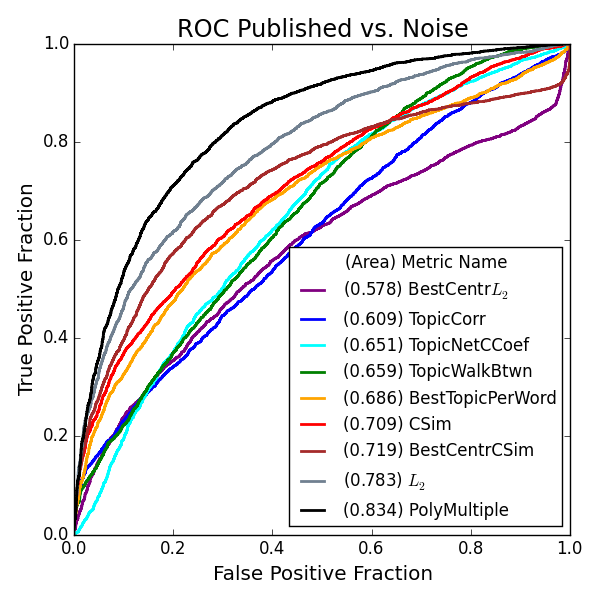}
\caption{\label{fig:published_roc}
The above ROC curves show the ability for each of our proposed methods to distinguish the \sysname results of published pairs from noise.
We use our system to generate hypotheses regarding 8,638 pairs, half from each set, on publicly available data released prior to 2,015.
We only show the best performing metrics from Section~\ref{sec:topicWalk} for clarity.
}
\end{minipage}\qquad
\begin{minipage}[b]{0.45\textwidth}
\centering
\includegraphics[width=0.8\linewidth]{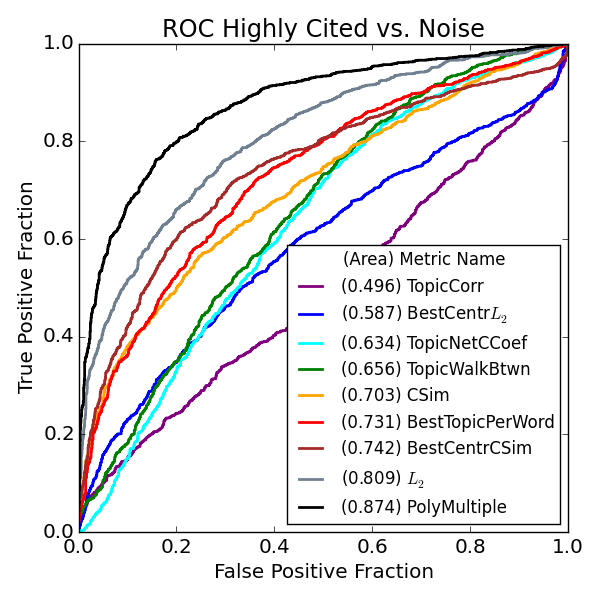}
\caption{\label{fig:highly_cited_roc}
The above ROC curves show the ability for each of our proposed methods to distinguish the \sysname results of highly-cited pairs from noise.
We identify 1,448 pairs who first occur in papers with over 100 citations published after our cut date.
To plot the above ROC curve, we also select an random subset of equal size from the noise pairs.
}
\end{minipage}
\end{figure*}

As described in Section~\ref{sec:validationChallenge}, our goal is to distinguish publishable connections from noise.
We run \sysname to generate topic models related to published, noise, and highly-cited pairs. Using this information, we plot ROC curves in Figures~\ref{fig:published_roc} and~\ref{fig:highly_cited_roc}, and summarize the results in Table~\ref{tab:results}.
These plots represent an analysis of 8,638 published vs. noise (PvN) pairs and 2,896 (HCvN) pairs (half of each set are noise). \emph{Unfortunately, no alternative general-purpose query HG systems that perform in a reasonable time are freely available for the comparison with our ranking methods.}

\noindent {\bf Topic Model Correlation} metric (see Section  \ref{sec:topicModelCorrelation}) is a poorly performing metric with an ROC area of 0.609 (PvN) and 0.496 (HCvN).
The core issue of this method is its sensitivity to the number of topics generated, and  given that we generate 100 topics per pair, we likely drive down performance through topics which are unrelated to the query.
In preliminary testing, we observe this intuition for queries with only 20 topics, but also find the network-biased metrics are less meaningful.
In Section~\ref{sec:deploymentChallenges} we overview a potential way to combine multiple topic models in our analysis.

Surprisingly, this metric is less able to distinguish highly-cited pairs, which we suppose is because highly-cited connections often bridge very distant concepts~\cite{rzhetsky2015choosing} and likely results in more noisy topic models.
Additionally, we may be able to limit this noise by tuning the number of topics returned from a query, as described in Section~\ref{sec:deploymentChallenges}.

\noindent {\bf $L_2$-based metrics} exhibit even more surprising results.
$\textsc{BestCentrL}_2$ performs poorly, with an ROC area of 0.578 (PvN) and 0.587 (HCvN), while the much simpler $L_2$ metric is exceptional, scoring a 0.783 (PvN) and 0.809 (HCvN).
We note that if two words are related, they are more likely to be closer together in our vector space.
We evaluate topic centroids based on their closeness to the midpoint between $a$ and $c$, normalized by the distance between them, so if that distance is small, the radius from the midpoint is small as well.
Therefore, it would seem that the distance between $a$ and $c$ is a better connection indication, and that the result of the centroid measurement is worse if this distance is small.

\noindent {\bf \csim-based metrics} are more straightforward.
The simple \csim metric scores a 0.709 (PvN) and 0.703 (HCvN), which is interestingly consistent given that the $L_2$ metric increases in ROC area given highly-cited pairs.
The \textsc{BestTopicPerWord} metric only scores a 0.686 (PvN), but increases substantially to 0.731 (HCvN).
The topic centroid method \textsc{BestCentroidCSim} is the best cosine-based metric with an ROC area of 0.719 (PvN) and 0.742 (HCvN).
This result is evidence that our initial hypothesis described in Figure~\ref{fig:topicsInVectorSpace} holds given cosine similarity, but as stated above, does not hold for Euclidean distance.

\noindent {\bf Topic network} metrics are all outperformed by simple $L_2$, but we see interesting properties from their results that help users to interpret generated hypotheses.
For instance, we see that \textsc{TopicWalkBtwn} is a negative indicator while \textsc{TopicWalkEigen} is positive.
Looking at the example in Figure~\ref{fig:topicGraph} we see that $a$ and $c$ are both far from the center of the network, connected to the rest of the topics through a very small number of high-betweenness nodes.
In contrast, we see that in the network created from a published pair, the path from $a$ to $c$ is more central.
We also see a denser clustering for the noise pair network, which is echoed by the fact that \textsc{TopicNetCCoef} and \textsc{TopicNetMod} are both negative indicators.
Lastly, we see that \textsc{TopicWalkLength} performs the best out of these network approaches, likely because it is most similar to the simple $L_2$ or \csim  ~metrics.

\noindent {\bf Combination of metrics,} $\textsc{PolyMultiple}$, significantly outperforms all others with ROC areas of 0.834 (PvN) and 0.874 (HCvN).
This is unsurprising because each other metric makes a different assumption about what sort of topic or vector configuration best indicates a published pair.
When each is combined, we see not only better performance, but their relative importances.
By studying the coefficients of our polynomial we observe that the two $L_2$-based metrics are most important, followed by the topic network methods, and finally by \textsc{TopicWalkCorr} and \textsc{BestTopicPerWord}.
Unsurprisingly, the coefficient signs correlate directly with whether each metric is a positive or negative indication as summarized in Table~\ref{tab:results}.
Additionally, the ordering of importance roughly follows the same ordering as the ROC areas.

\begin{table}
\centering
\begin{tabular}{|l|c|c|}
\hline
Metric Name & PvN ROC & HCvN ROC \\
\hline
\textsc{PolyMultiple} & 0.834 & 0.874\\
$L_2$* & 0.783 & 0.809 \\
\csim & 0.709 & 0.703 \\
\textsc{BestCenterL}$_2$ & 0.578 & 0.587 \\
\textsc{BestCenterCSim} & 0.719 & 0.742 \\
\textsc{BestTopicPerWord} & 0.686 & 0.731 \\
\textsc{TopicCorr} & 0.609 & 0.496 \\
\textsc{TopicWalkLength}* & 0.740 & 0.778 \\
\textsc{TopicWalkBtwn}* & 0.659 & 0.658 \\
\textsc{TopicWalkEigen} & 0.585 & 0.582 \\
\textsc{TopicNetCCoef}* & 0.651 & 0.638 \\
\textsc{TopicNetMod}* & 0.659 & 0.628 \\
\hline
\end{tabular}
\vspace{0.1in}
\caption{The above summarizes all ROC area results for all considered metrics on the set of published vs. noise pairs (PvN) and highly-cited vs. noise pairs (HCvN). 
Metrics marked with a (*) have been sorted in reverse order for the ROC calculations.
}\label{tab:results}
\end{table}

\section{Case-Study: HAND and DDX3 Candidate Selection}\label{sec:applied_results}

Our proposed validation method is rooted in the process of candidate selection.
To demonstrate our method's applicability to real-world scenarios, we applied the above methods to a series of queries surrounding Human Immunodeficiency Virus -associated dementia (or HIV-associated neurodegenerative disease, HAND).
HAND is one of the most common and clinically important complications of HIV infection~\cite{kovalevich2012neuronal}.
The brain-specific effects of HIV are of great concern because the HIV-infected population is aging and unfortunately revealing new pathologies~\cite{spudich2013hiv, bilgrami2014neurologic}.
About 50\% of HIV-infected patients are at risk of developing HAND, which might be severely worsened by abusing drugs such as cocaine, opioids and amphetamines~\cite{beyrer2010epidemiologic, buch2012cocaine}.

We generated over 30,000 queries, each between HAND and a gene from the HUGO Gene Nomenclature Committee dataset~\cite{hgnc2017}.
The network that generated these results consisted of the 2017 \medline dataset, the 2017AB \umls release, and the 2016 \semmeddb release (latest at the time).
We trained \ftext using all of the available titles and abstracts, about 27 million in total, and selected a dimensionality of 500 for our word embeddings.
Our results consist of each disease-gene query ranked by our \textsc{PolyMultiple} metric.

Based on this ranking we select the first {\textasciitilde}1000 genes for further analysis.
We observe that many of the top genes --- such as APOE-4, T-TAU, and BASE1, which occur in our top five --- are known to be linked to dementia.
So to direct our search to yet-unknown connections, we select those genes that have no previous connection to HAND, but still ranked highly overall.
This process limits our search to those proteins that have known selective compounds, which were often tested animal models or clinical trials.

\begin{figure}
\centering
\includegraphics[width=0.9\linewidth]{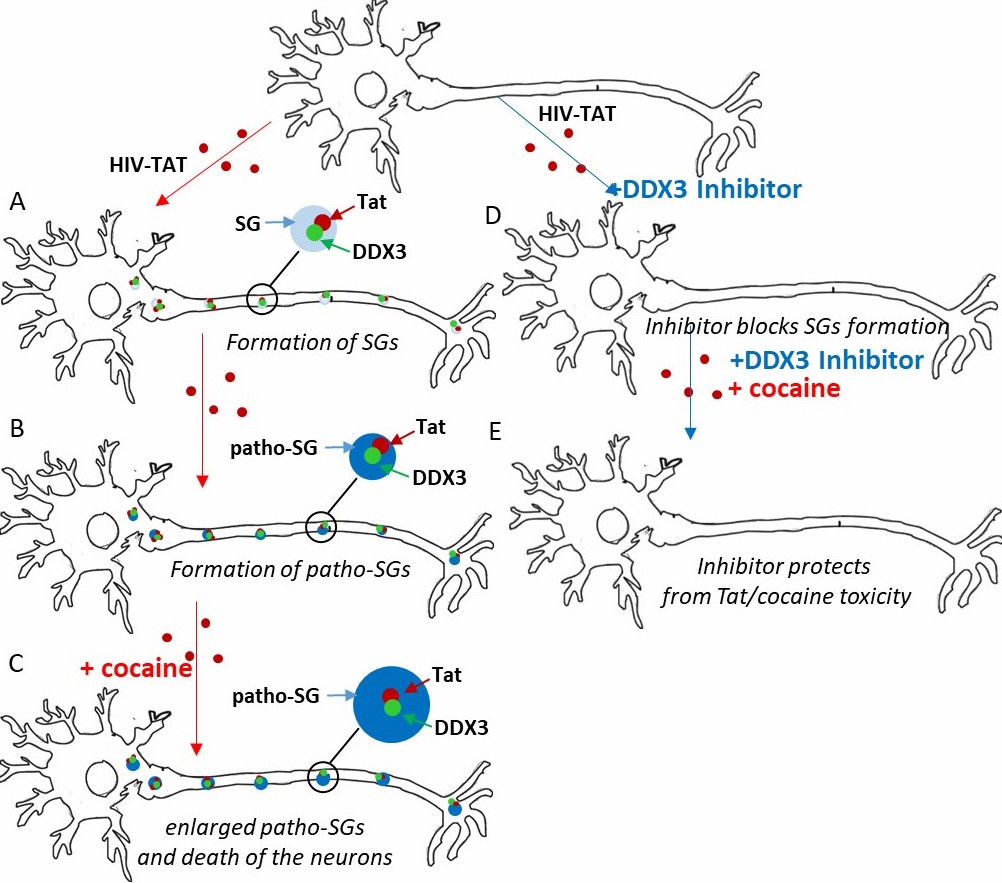}
\caption{\label{fig:med_res}
Scheme of the hypothesis of Stress-Granule dependent mechanism of neuroprotection by DDX3 inhibitor. Neurons are curved figures. Treatment with HIV-Tat leads to DDX3-dependent formation of SGs (A), which transform from ``normal'' to ``pathological'' (B). The addition of cocaine further enlarges the SGs and leads to the death of the neurons (C). Treatment with DDX3 specific inhibitor blocks DDX3 enzymatic activity and Tat-dependent SG formation (D) and protects the neurons from cocaine-induced death (E).
}
\end{figure}

From this candidate set we selected Dead Box RNA Helicase 3 (DDX3).
We tested the activity of a DDX3 inhibitor on the tissue culture model of HAND, which is widely used for the analysis combine neurotoxicity of HIV proteins and drugs of abuse.
Here we tested the effect of the DDX3 inhibition on combined toxicity of most toxic HIV protein, Trans-Activator of Transcription (Tat).
The mouse cortical neurons had been treated with HIV Tat followed by the addition of cocaine.
The combination of Tat and cocaine kills more than 70\% of the neurons, while the inhibitor protects the neurons from Tat/cocaine toxicity (Figure~\ref{fig:med_res}). 

Based on the analysis, we formulate following hypothesis:
Exposing neurons with Tat protein causes internal stress and results in the formation of Stress-Granules (SGs) --- the structures in cytoplasm formed by multiple RNAs and proteins.
These gel-like structures sequester cellular RNA from translation, and the formation of SGs requires enzymatically active Dead Box RNA Helicase 3.
The formation of SGs also allows the neurons to wait out the stress.
However, prolonged stress associated with HIV-Tat treatment leads to the formation of pathological stress granules, which are denser and have a different composition relative to ``normal'' ones.
Additional exposure to cocaine further exaggerates the ``pathological'' SGs and eventually causes neuronal death.
The hypothesis, initially generated with \sysname, led to the following finding: \emph{Treatment with a DDX3-specific inhibitor blocks the enzymatic activity of the DDX3.
This lack of enzymatic activity, in turn, blocks Tat-dependent stress granules from formating and protects neurons from the combined toxicity of Tat and cocaine.}
In Figure \ref{fig:med_res}, we demonstrate the hypothesis scheme.   
Thus, the application of the automated HG system pointed to a new avenue for anti-HAND therapy and to the prototype of a small molecule for drug development. 

\section{Related Work and Proposed Validation}\label{sec:relatedWork}

The  HG community struggles to validate its systems in a number of ways.
Yetisgen-Yildiz and Pratt, in their chapter ``Evaluation of Literature-Based Discovery Systems,'' outline four such methods (M1-M4)~\cite{bruza2008literature, yetisgen2008evaluation}.

\noindent {\bf M1: Replicate Swanson's Experiments.} Swanson, during his development of ARROWSMITH~\cite{smalheiser1998using}, worked alongside medical researchers to uncover a number of new connections.
These connections include the link between Raynaud's Disease and Fish Oil~\cite{swanson1986fish}, the link between Alzheimer's Disease and Estrogen~\cite{smalheiser1996linking} and the link between Migraine and Magnesium~\cite{swanson1988migraine}.
As discussed in~\cite{yetisgen2008evaluation}, a number of projects have centered their validation effort around Swanson's results~\cite{heo2014inferring, blake2002automatically, srinivasan2004text, hu2005semantic, pratt2003litlinker}.
These efforts always rediscover a number of findings using information before Swanson's discovery date, and occasionally apply additional metrics such as precision and recall in order to quantify their results~\cite{han2011data}.

While limiting discussion to Swanson's discoveries reduces the domain of discovery drastically, at its core this method builds confidence in a new system through its ability to find known connections.
We expand on this idea by validating automatically and on a massive scale, freeing our discourse from a single researcher's findings.

\noindent {\bf M2: Statistical Evaluation.} 
Hristovski et al. validate their system by studying a number of relationships and note their confidence and support with respect to the \medline document set~\cite{hristovski2005using}.
Then, they can generate potential relationships for the set of new connections added to \umls~\cite{MedicineUS2009} or OMIM~\cite{hamosh2005online}.
By limiting their method to association rules, Hristovski et al. note that they can validate their system by predicting UMLS connections using data available prior to their publications.
Therefore, this method is similar to our own, but we notice that restricting discussion to only \umls gene-disease connections results in a much smaller set than the predicate information present with \semmeddb.

Pratt et al. provide additional statistical validation for their system LitLinker~\cite{pratt2003litlinker}.
This method also calculates precision and recall, but this time focusing on their $B$-set of returned results.
Their system, like ARROWSMITH~\cite{smalheiser1998using}, returns a set of intermediate terms which may connect two queried entities.
Pratt et al. run LitLinker for a number of diseases on which they establish a set of ``gold standard'' terms.
Their method is validated based on its ability to list those gold-standard terms within its resulting $B$-sets.
\emph{This approach requires careful selection of a (typically small) set of gold-standard terms, and is limited to ``ABC'' systems like ARROWSMITH, which are designed to identify term lists}~\cite{smalheiser2012literature}.

\noindent {\bf M3: Incorporating Expert Opinion.} 
This ranges from comparisons between system output and expert output, such as the analysis done on the Manjal system~\cite{srinivasan2004text}, to incorporating expert opinion into gold-standard terms for LitLinker~\cite{pratt2003litlinker}, to running actual experiments on potential results by Wren et al.~\cite{wren2004knowledge}.
Expert opinion is at the heart of many recent systems~\cite{wang2011finding, sybrandt2017moliere, liu2014diseaseconnect, swanson1986undiscovered}, including the previous version of our own.
This process is both time consuming and risks introducing significant bias into the validation.

Spangler incorporates expert knowledge in a more sophisticated manner through the use of visualizations~\cite{spangler2015accelerating,spangler2014automated}.
This approach centers around visual networks and ontologies produced automatically, which allows experts to see potential new connections as they relate to previously established information.
This view is shared by systems such as DiseaseConnect~\cite{liu2014diseaseconnect} which generates sub-networks of ONIM and GWAS related to specific queries.
Although these visualizations allow users to quickly understand query results, they do not lend themselves to a numeric and massive evaluation of system performance.

BioCreative, a set of challenges focused on assessing biomedical text mining, is the largest endeavor of its kind, to the best of our knowledge~\cite{Hirschman2005}.
Each challenge centers around a specific task, such as mining chemical-protein interactions, algorithmically identifying medical terms, and constructing causal networks from raw text.
Although these challenges are both useful and important, their tasks fall under the umbrella of \emph{information retrieval} (and not HG) because their tasks compare expert analysis with software results given the same text.

\noindent {\bf M4: Publishing in the Medical Domain.} This method is exceptionally rare and expensive.
The idea is to take prevalent potential findings and pose them to the medical research community for another group to attempt.
Swanson and Smalheiser rely on this technique to solidify many of their early results, such as that between magnesium deficiency and neurologic disease~\cite{smalheiser1994assessing}.

Bakkar et al. take a similar approach in order to demonstrate the efficacy of Watson for Drug Discovery~\cite{bakkar2018artificial,spangler2014automated}
To do so, this work begins by identifying 11 RNA-binding proteins (RBPs) known to be connected to Amyotrophic Lateral Aclerosis (ALS).
Then, the automated system uses a recommender system to select RPBs that exhibit similar connection patterns within a large document co-occurrence network.
Domain scientists then explore a set of candidates selected by the computer system, and uncover five RPBs that were previously unrelated to ALS.

An alternative to the domain-scientist approach is taken by Soldatova and Rzhetsky wherein a ``robot scientist'' automatically runs experiments posed by their  HG system~\cite{Soldatova2011,rzhetsky2016big}.
This system uses logical statements to represent their hypotheses, so new ideas can be posed through a series of implications.
Going further, their system even identifies statements that would be the most valuable if proven true~\cite{rzhetsky2015choosing}.
\emph{However, the scope of experiments that a robot scientist can undertake is limited; in their initial paper, the robot researcher is limited to small-scale yeast experiments.
Additionally, many groups cannot afford the space and expense that an automated lab requires.}

\section{Deployment Challenges and Open Problems} \label{sec:deploymentChallenges}

\noindent\textbf{Validation Size.} Our proposed validation challenge involves ranking millions of published and noise query pairs.
However, in Section~\ref{sec:results} we show our results on a randomly sampled subset of our overall challenge set.
This was necessary due to performance limitations of MOLIERE, a system which initially required a substantial amount of time and memory to process even a single hypothesis.
To compute these results, we ran 100 instances of \sysname, each on a 16 core,  64 GB RAM machine connected to a ZFS storage system.
Unfortunately, performance limitations within ZFS created a bottleneck that both limited our results and drastically reduced cluster performance overall.
Thus, our results represent a set of predicates that we evaluated in a limited time period.

\noindent\textbf{System Optimizations.}
While performing a keyword search, most network-centered systems are either I/O or memory bound simply because they must load and traverse large networks.
In the case of \sysname, we initially spent hours trying to find shortest paths or nearby abstracts.
But, we found a way to leverage our embedding space and our parallel file system in order to drastically improve query performance.
In brief, one can discover a relevant knowledge-network region by inducing a subnetwork on $a$ and $c$ and expanding that selection by adding $i^{th}$ order neighbors until $a$ and $c$ are connected.
From our experiments, $i$ rarely exceeds 4.
This increases performance because, given a parallel file system and $p$ processors, identifying the subnetwork from an edge list file is in order $\mathcal{O}(ni/p)$.
The overall effect reduced the wall-clock runtime of a single query from about 12 hours to about 5-7 minutes.
Additionally, we reduced the memory requirement for a single query from over 400GB to under 16GB.


\noindent\textbf{Highly-Cited Predicates.}
Identifying highly-cited predicates requires that we synthesize information across multiple data sources.
Although \semmeddb contains \medline references for each predicate, neither contains citation information.
For this, we turn to Semantic Scholar because not only do they track citations of medical papers, but they allow a free bulk download of metadata information (many other potential sources either provide a very limited API or none at all).
In order to match Semantic Scholar data to \medline citation, it is enough to match titles.
This process allows us to get citation information for many \medline documents, which in turn allow us to select predicates whose first occurrence was in highly-cited papers.
We explored a number of thresholds for what constitutes ``highly cited'' and selected 100 because it was a round number and selected a sizable predicate set.
Because paper citations follow a power-law distribution, any change drastically effects the size of this set.
We note that the set of selected predicates was also limited by the quality of data in Semantic Scholar, and that the number of citations identified this was appeared to be substantially lower than that reported by other methods.



\noindent\textbf{Quality of Predicates.}
Through our above methods we learned that careful ranking methods can distinguish between published and noise predicates, but there is a potential inadequacy in this method.
Potentially, some predicates that occur within our published say may be untrue.
Additionally, it is possible that a noise predicate may be discovered to be true in the future.
If \sysname ranks the published predicate which is untrue below the noise predicate which is, the result would be a lower ROC area.
This same phenomena is addressed by Yetisgen-Yildiz and Pratt when they discuss the challenges present in validating literature-based discovery systems~\cite{yetisgen2008evaluation} --- if a HG systems goal is to identify novel findings, then it \emph{should} find different connections than human researchers.

We show through our results that despite an uncertain validation set, there are clearly core differences between publishable results and noise, which are evident at scale.
Although there may be some false positives or negatives, we see through our meaningful ROC curves that they are far outnumbered by more standard predicates.

\noindent\textbf{Comparison with ABC Systems.}
Additionally, we would like to explore how our ranking methods apply to traditional ABC systems.
Although there are clear limitations to these systems~\cite{smalheiser2012literature}, many of the original systems such as ARROWSMITH follow the ABC pattern.
These systems typically output a list of target terms and linking terms, which could be thought of as a topic.
If we were to take a pre-trained embedding space, and treated a set of target terms like a topic, we could likely use our methods from Section~\ref{sec:methodology} to validate any ABC system.

\noindent\textbf{Verb Prediction.}
We noticed, while processing \semmeddb predicates, that we can improve \sysname if we utilize verbs.
\semmeddb provides a handful of verb types, such as ``TREATS,'' ``CAUSES,'' or ``INTERACTS\_WITH,'' that suggest a concrete relationship between the subject and object of a sentence.
\sysname currently outputs a topic model that can be interpreted using our new metrics, but does not directly state what sort of connection may exist between $a$ and $c$.
Thus we would like to explore accurately predicting these verb types given only topic model information.

\noindent {\bf Interpretability of Hypotheses} remains one of the major problems in HG systems. Although topic-driven HG partially resolve this issue by producing readable output, we still observe many topic models $T$ (i.e., hypotheses) whose $T_i\in T$ are not intuitively connected with each other. While the proposed ranking is definitely helpful for understanding $T$, it still does not fully resolve the interpretability problem. One of our current research directions is to tackle it using text summarization techniques.

\noindent {\bf Scope.}
While we focus on biomedical science, any field that is accurately described by \emph{entities} that \emph{act} on one another benefits from our network and text mining methods.
For instance, economic entities, such as governments or the upper/lower class, interact via actions such as regulation or boycott.
Similarly, patent law consists of inventions and the components that comprise them.
Mathematics, in contrast, is not served by this representation --- the algebra does not \emph{act} on other math entities. 
Here automatic theorem proving is better equipped to generate hypotheses.
We are presently unsure if the same is true for computer science.





\bibliographystyle{IEEEtran}
%
\bibliography{bibFile.bib}

\end{document}